\documentclass{JHEP3}
\usepackage{amsmath}
\usepackage{amssymb}
\usepackage{graphicx}
\usepackage{cite}
\usepackage{tabularx}
\title{Large logarithmic rescaling of the scalar condensate: a subtlety
with substantial phenomenological implications}

\author{P. Cea \\ Dipartimento
Interateneo di Fisica, Universit\`a di Bari  and INFN - Sezione di Bari, \\
I-70126 Bari, Italy \\
E-mail:  \email{Paolo.Cea@ba.infn.it} }

\author{M. Consoli \\ INFN - Sezione di Catania, I-95123 Catania,
Italy \\
E-mail:  \email{Maurizio.Consoli@ct.infn.it} }

\author{L. Cosmai \\ INFN - Sezione di  Bari, I-70126 Bari,
Italy \\
E-mail:  \email{Leonardo.Cosmai@ba.infn.it} }

\received{\today}

\abstract{Lattice data, taken since 1998
near the critical line of a 4D Ising model, have been supporting
the large logarithmic rescaling of the scalar condensate
predicted in the alternative description of symmetry breaking proposed
by Consoli
and Stevenson. This conclusion has been challenged in a recent paper by
Balog et al. In this paper we respond to the criticism of these authors,
recapitulate the theoretical and numerical
evidences in favour of the alternative interpretation
of `triviality' and reiterate our conclusion: `triviality', by itself,
cannot be used to place upper bounds on the Higgs boson mass.}


\preprint{BARI-TH 504/2005}

\begin{document}

\section{Introduction}
\label{Introduction}

Recent lattice data
\cite{Cea:2004ka}, collected near the critical line of
a 4D Ising model, support
the large logarithmic rescaling of the scalar condensate predicted
in an alternative description of symmetry breaking in $\Phi^4$ theories,
see
Refs.\cite{Consoli:1994jr,Consoli:1997ra,Consoli:1999ni}.
This result, while confirming previous numerical indications obtained in
Refs.~\cite{Cea:1998hy,Cea:1999kn,Cea:1999zu}, would have a substantial
phenomenological implication:
one cannot use `triviality' to place upper bounds on the Higgs boson mass.

This point of view has been challenged in a recent paper
\cite{Balog:2004zd} by Balog et
al.. These authors, referring just to
Ref.\cite{Cea:2004ka}, while otherwise
ignoring the previous numerical indications of
Refs.~\cite{Cea:1998hy,Cea:1999kn,Cea:1999zu},
draw the opposite conclusion: the standard
interpretation, as they say the `Conventional Wisdom' (CW), is completely
consistent
with all lattice data. The aim of this paper is to
 respond to their
criticism, recapitulate in a unified framework the results of
Ref.\cite{Cea:2004ka} and
Refs.~\cite{Cea:1998hy,Cea:1999kn,Cea:1999zu}, and reiterate our
conclusion:
`triviality', by itself, cannot
be used to place upper bounds on the Higgs boson mass.

\section{The rescaling of the scalar condensate}
\label{The rescaling}

Before entering the details of the controversy, we shall first remind once
again in this section why
in a spontaneously broken phase there are {\em two} basically
different definitions of the field rescaling. In fact,
the widespread
skepticism concerning the interpretation of `triviality' proposed
in Refs.\cite{Consoli:1994jr,Consoli:1997ra,Consoli:1999ni} originates
from a non sufficient appreciation of this crucial point.

To this end, let us introduce
the bare `lattice' field $\Phi_B(x)=\Phi_{\text{latt}}(x)$
(i.e. as defined at a locality scale fixed by the ultraviolet
cutoff $\Lambda\sim \pi/a$, $a$ being the lattice spacing)
and its expectation value (the `scalar condensate')
\begin{equation}
\label{vB}
v_B=\langle\Phi_B\rangle
\end{equation}
Connecting to the stability analysis, the values $\pm v_B$
represent the absolute minima
of the effective potential $V_{\text{eff}}(\varphi_B)$ of the theory.
We shall also introduce the bare shifted fluctuation field
\begin{equation}
\label{hB}
h_B(x)=\Phi_B(x)-
\langle\Phi_B\rangle,
\end{equation}
whose expectation value vanishes by definition.

Now, a first natural definition of the field rescaling, say
$Z=Z_{\text{prop}}$, is obtained from
the residue of the shifted-field propagator
\begin{equation}
\label{pole}
    G_{\text{pole}}(s)\sim \frac{Z_{\text{prop}}}{ s - m^2_H}
\end{equation}
near the physical mass-shell $s=m^2_H$. This
can be used to define a renormalized
fluctuation field $h_R(x)$
\begin{equation}
\label{hr}
h_B(x)=\sqrt{Z_{\text{prop}}} h_R(x)
\end{equation}
whose propagator has the canonical form.

This first definition of the field rescaling is
constrained by the
K\'allen-Lehmann representation
to lie in the range
$0 < Z_{\text{prop}}\leq 1$, the free-field limit corresponding to the case
$Z_{\text{prop}} = 1$. This can be rigorously established for the local,
Lorentz-covariant theory. If this is viewed as the continuum limit
of the cutoff theory, one expects \cite{Zimmermann:1970aa}
\begin{equation}
\label{zimmermann} Z^{-1}_{\text{prop}}=1+\int^{\Lambda^2}_{s_o}ds\rho_\Lambda(s)
\end{equation}
where $s_o$ denotes the continuum threshold
($s_o=(2m_H)^2$ in perturbation theory) and
$\rho_\Lambda(s)$ the spectral function. For this reason, in the continuum
limit $\Lambda \to \infty$, where
according to `triviality' the spectral function
 should tend to $\delta(s-m^2_H)$,
$Z_{\text{prop}}$ should tend to unity.

Another definition of the field rescaling, say $Z\equiv Z_\varphi$,
is peculiar of a broken-symmetry phase. It indicates
the rescaling that is needed to relate the {\it physical} vacuum field
$v_R$ to the bare $v_B$, i.e.
\begin{equation}
\label{vR}
v_R= \frac{v_B}{\sqrt{Z_\varphi}}  \,.
\end{equation}
By {\em physical}, we mean that the second derivative
of the effective potential
$V''_{\text{eff}}(\varphi_R)$
evaluated at $\varphi_R=\pm v_R$, is precisely given by the physical Higgs
boson mass squared, i.e.
\begin{equation}
V''_{\text{eff}}(v_R)= m^2_H
\end{equation}
 This is very simple to understand.
$V''_{\text{eff}}(\varphi_R)$ represents the renormalized
2-point function evaluated at an
external 4-momentum $p_{\mu}=0$. For $\varphi_R=v_R$,
this should match the inverse of the renormalized connected
propagator at zero momentum. In a `trivial' theory, in the continuum limit,
this has the simple free-field form $G_R(0)= 1/m^2_H$.

Therefore, this other definition is equivalent to the relation
\begin{equation}
\label{z1phi}
      Z_\varphi= \frac{
V''_{\text{eff}}(v_R)}
{V''_{\text{eff}}(v_B)}=m^2_H \chi_2 (0)
\end{equation}
where $\chi_2(0)=1/V''_{\text{eff}}(v_B)$
is the bare zero-momentum susceptibility.

Let us now explain why $Z_{\text{prop}}$ and
      $Z_\varphi$ are completely different
physical quantities. To this end, we shall
consider the class of `triviality compatible', gaussian-like
approximations to the effective potential, say
$V_{\text{eff}}(\varphi_B)\equiv
V_{\text{triv}}(\varphi_B)$, where the shifted fluctuating field
is governed by an effective quadratic hamiltonian. In this class, either
$Z_{\text{prop}} = 1$ identically or it tends to unity
in the continuum limit
$\Lambda \to \infty$. In spite of this, the rescaling of the vacuum field
\begin{equation}
\label{z2phi}
      Z_\varphi=  m^2_H \chi_2 (0) \sim \ln \Lambda
\end{equation}
diverges logarithmically.

This result is easily recovered noticing that the
class $V_{\text{triv}}$ includes the one-loop
potential, the gaussian approximation and the infinite set of
post-gaussian calculations where the effective potential
reduces to the sum of a classical background energy and of the zero-point
energy of the {\em free}
massive shifted field with a $\varphi_B-$dependent mass,
$\Omega=\Omega(\varphi^2_B)$. For instance, at one loop one finds
$\Omega^2(\varphi^2_B)=m^2_B+\frac{\lambda_B\varphi^2_B}{2}$, $m^2_B$ and
$\lambda_B$ denoting respectively the bare mass squared and the bare
fourth-order self-coupling. In gaussian and post-gaussian calculations, the
functional dependence
$\Omega=\Omega(\varphi^2_B)$ is determined upon a minimization procedure of
the energy functional in a suitable class of quantum trial states.
In all cases, the basic mass
parameter of the broken-symmetry phase is obtained through the relation
$m_H=\Omega(v^2_B)$.

Therefore, the various approximations belonging to
$V_{\text{triv}}$ share
the same simple general structure, up to non-leading terms that vanish faster
when $\Lambda \to \infty$
(see Ref.\cite{Consoli:1999ni} for the general
argument, Ref.\cite{Branchina:1993rj} for the
gaussian approximation and
Ref.\cite{Ritschel:1994vr} for the case of a
post-gaussian approximation). For the particularly simple case of the
classically
scale-invariant theory (the `Coleman-Weinberg regime') this is given by
\begin{equation}
\label{vtriv}
V_{\text{triv}}(\varphi_B)=
\frac{ \lambda_{\text{eff}} \varphi^4_B}{4!} +
\frac{\Omega^4(\varphi^2_B) }{64\pi^2}
(\ln\frac{\Omega^2(\varphi^2_B)}{\Lambda^2} -\frac{1}{2})
\end{equation}
with
\begin{equation}
\label{omega}
\Omega^2(\varphi_B)=
\frac{\lambda_{\text{eff}} \varphi^2_B}{2}
\end{equation}
all differences among the various approximations being isolated in the
relation between the effective coupling
$\lambda_{\text{eff}}=\lambda_{\text{eff}}(\Omega)$ and the bare coupling.

For instance, at one loop
$\lambda_{\text{eff}}=\lambda_B$ while in the gaussian approximation
(see Ref.\cite{Branchina:1993rj})
$\lambda_{\text{eff}}$ corresponds to resum all one-loop bubbles with mass
$\Omega=\Omega(\varphi^2_B)$ so that
\begin{equation}
\label{leff}
\lambda_{\text{eff}}(\Omega)=\frac{\lambda_B}
{1+ \frac{\lambda_B}{16\pi^2} \ln\frac{\Lambda}{\Omega} }
\end{equation}
Now, since
all approximations display the same general structure in Eqs.(\ref{vtriv}) and
(\ref{omega}), upon minimization and
after setting $\varphi_B=\pm v_B$, one finds
the same leading logarithmic trend of the effective coupling
\begin{equation}
\label{ltriv}
 \lambda_{\text{eff}}(m_H) \sim
\frac{ 16\pi^2}{ 3 \ln \frac{\Lambda}{m_H}}
\end{equation}
Formally, this is precisely
the same relation obtained in perturbation theory
from the leading-order (LO) renormalized coupling
\begin{equation}
\label{lambdaR}
 \lambda_{\text{LO}}(m_H)
=\frac{\lambda_B}{1+
\frac{3\lambda_B}{ 16\pi^2}\ln \frac{\Lambda}{m_H} }
\end{equation}
after sending $\lambda_B \to \infty$ at the Landau pole
\begin{equation}
\label{lambdaR2}
 \lambda_{\text{LO}}(m_H) \sim
\frac{ 16\pi^2}{ 3 \ln \frac{\Lambda}{m_H}}
\end{equation}
Therefore, one might be tempted to conclude that, at least at
the leading logarithmic level, the alternative
picture of Refs.\cite{Consoli:1994jr,Consoli:1997ra,Consoli:1999ni}, is
equivalent to the CW.

However, there is one notable difference: the quadratic shape of the
`trivial' effective potential Eq.(\ref{vtriv}) at its absolute minima,
namely the inverse susceptibility $V''_{\text{triv}}(v_B)$, is
{\em not} given by $m^2_H=\Omega^2(v^2_B)$. Rather, one finds
\begin{equation}
\label{shape}
V''_{\text{triv}}(v_B)\sim
 \lambda_{\text{eff}} m^2_H \sim \frac{m^2_H}{ \ln \Lambda }
\end{equation}
thus leading to Eq.(\ref{z2phi}).

Eq.~(\ref{z2phi}) has
substantial phenomenological implications.
In fact, using Eq.~(\ref{vR}) with a
logarithmically divergent $Z_\varphi$ as in Eq.~(\ref{z2phi}), one finds
\begin{equation}
\label{vrvb}
v^2_R \sim \frac{ v^2_B }{ \ln \Lambda }
\end{equation}
instead of the conventional relation obtained for a trivial unit rescaling
\begin{equation}
\label{vrvbpt}
\left[ v^2_R \right]_{\text{CW}} \sim v^2_B
\end{equation}
Now, whatever the $v_R-v_B$ relation, the Higgs boson mass squared goes like
\begin{equation}
\label{stable}
m^2_H  \sim  \frac{ v^2_B }{ \ln \Lambda }
\end{equation}
This becomes
\begin{equation}
\label{usual}
m^2_H \sim \frac{ \left[ v^2_R \right]_{\text{CW}} }{ \ln \Lambda }
\end{equation}
by using Eq.(\ref{vrvbpt}) but becomes
\begin{equation}
\label{golden}
                     m^2_H \sim v^2_R
\end{equation}
if one defines $v_R$ through Eq.(\ref{vrvb}). In the latter case,
$m_H$ and $v_R$ scale uniformly in the continuum limit.

Therefore, assuming to relate the $v_R$ of Eq.(\ref{vrvb})
(and {\it not} the conventional
$\left[ v_R \right]_{\text{CW}} \sim v_B$ of Eq.(\ref{vrvbpt}))
to some physical scale (say 246 GeV),
a measurement of $m_H$ would not provide any information on the magnitude
of $\Lambda$ since, according to Eq.(\ref{golden}), the ratio
$C=m_H/v_R$ is now a cutoff-independent quantity. Moreover,
in this approach, the quantity $C$ does not represent
the measure of any {\em observable} interaction (see the Conclusions
of Ref.\cite{Agodi:1995qv}).

We emphasize that the difference between $Z_\varphi$ and $Z_{\text{prop}}$
has a precise physical meaning being a
distinctive feature of the
Bose condensation phenomenon~\cite{Consoli:1999ni}.
In the class of `triviality-compatible'
approximations to the effective potential,
one finds
$C=m_H/v_R=2 \pi \sqrt{2 \zeta}$, with $0< \zeta \leq 2$~\cite{Consoli:1999ni},
$\zeta$
being a cutoff-independent number determined by the quadratic shape of
the effective potential
$V_{\text {eff}}(\varphi_R)$
at $\varphi_R=0$. For instance, $\zeta=1$
corresponds to the classically scale-invariant case or `Coleman-Weinberg
regime'.

As for the standard interpretation of `triviality', the
gaussian effective potential approach can
also be extended to any number N of scalar field components
\cite{Stevenson:1987nb}.
In particular, when studying the continuum limit in the
large-N limit of the theory, one has
to take into account the non-uniformity of the two limits,
cutoff $\Lambda \to \infty$ and $N \to \infty$
\cite{Ritschel:1992ss,Ritschel:1994vr}. This is crucial
to understand the difference with respect to the standard large-N
analysis.

\section{Lattice tests of the alternative interpretation of `triviality'.
Part 1.}
\label{latticetest1}

To test the alternative picture of `triviality' of Refs.
\cite{Consoli:1994jr,Consoli:1997ra,Consoli:1999ni}
against the CW
one can run numerical simulations of the theory and
check the scaling properties of the squared Higgs lattice mass
against those of the inverse
zero-momentum lattice susceptibility.

The numerical simulations of Ref.\cite{Cea:2004ka}
(and of Refs.\cite{Cea:1998hy,Cea:1999kn,Cea:1999zu})
were performed in the Ising limit
that traditionally has been chosen as a convenient laboratory for the
numerical analysis of the theory.

In this limit,
a one-component $\Phi^4_4$ theory becomes governed by the lattice
action
\begin{equation}
\label{ising}
S_{\text{Ising}} = -\kappa \sum_x\sum_{\mu} \left[ \phi(x+\hat
e_{\mu})\phi(x) + \phi(x-\hat e_{\mu})\phi(x) \right]
\end{equation}
where $\phi(x)$ takes only the values $\pm 1$.

Using the Swendsen-Wang \cite{Swendsen:1987ce},
and Wolff~\cite{Wolff:1989uh} cluster algorithms we computed
the bare magnetization:
\begin{equation}
\label{baremagn}
 v_B=\langle |\phi| \rangle \quad , \quad \phi \equiv \sum_x
\phi(x)/L^4
\end{equation}
(where $\phi$  is the average field for each lattice configuration) and
the bare zero-momentum susceptibility:
\begin{equation}
\label{chi}
 \chi_{\text{latt}}=L^4 \left[ \left\langle |\phi|^2
\right\rangle - \left\langle |\phi| \right\rangle^2 \right] .
\end{equation}
We report in Table 1 our determinations
of $v_B$ and  $\chi_{\text{latt}}$. Other values,
obtained over the years by different authors, are reported
in Table 1 of Ref.~\cite{Balog:2004zd}.

Checks of the logarithmic trend predicted in
Eq.(\ref{z2phi}) can be performed using different methods.
In a first indirect approach, discussed in this Section, one can use
the fact that, both in perturbation theory and according to
Refs.~\cite{Consoli:1994jr,Consoli:1997ra,Consoli:1999ni}, the bare
field expectation value $v_B$ is predicted to diverge logarithmically
in units of the physical Higgs boson mass $m_H$, i.e.
\begin{equation}
\label{v2Blog}
\frac{v^2_B}{m^2_H} \sim  |\ln(\kappa-\kappa_c)| \,.
\end{equation}
Therefore, following CW, where
the zero-momentum susceptibility
is predicted to scale uniformly with the inverse squared Higgs mass
\begin{equation}
\label{mrchitau}
      \left[ m^2_H \chi_2(0) \right]_{\text{CW}}\sim 1
\end{equation}
one expects
\begin{equation}
\label{v2Bchi2pt}
      \left[ v^2_B \chi_2(0) \right]_{\text{CW}}\sim
|\ln(\kappa-\kappa_c)|
\end{equation}
On the other hand, in the approach of
Refs.~\cite{Consoli:1994jr,Consoli:1997ra,Consoli:1999ni} one
predicts
$Z_\varphi=\chi_2(0)m^2_H \sim \ln(\Lambda)$ so that, in this case,
one would rather expect (CS=Consoli-Stevenson)
\begin{equation}
\label{v2Bchi2cs}
      \left[ v^2_B \chi_2(0) \right]_{\text{CS}}\sim |\ln(\kappa-\kappa_c)|^2
\end{equation}
The two leading-order predictions in
Eq.~(\ref{v2Bchi2pt}) and in Eq.~(\ref{v2Bchi2cs}) were directly
compared with the lattice data for the product $v^2_B \chi_2(0)$ reported
in Table 3 of Ref.\cite{Cea:2004ka}.
These data were fitted to a 3-parameter form
\begin{equation}
\label{Ffit}
\alpha |\ln(\kappa-\kappa_c)|^\gamma
\end{equation}
where $\alpha$ is a normalization constant and one can set the
exponent $\gamma=1$, according to
Eq.~(\ref{v2Bchi2pt}), or $\gamma=2$ according
to Eq.~(\ref{v2Bchi2cs}). Using this type of functional form
the results of the fit to the lattice data
single out unambiguously the alternative picture of
Refs.~\cite{Consoli:1994jr,Consoli:1997ra,Consoli:1999ni}, i.e.
$\gamma=2$, over the value $\gamma=1$
(see Fig.2 of Ref.\cite{Cea:2004ka}).

However, it has been pointed out by the authors of
Ref.\cite{Balog:2004zd} that the lattice data
can also be reproduced for $\gamma=1$, once the scale within
the log is left as a free parameter, thus effectively simulating the presence
of next-to-leading corrections. Indeed, in this case
one gets a good fit in both cases with precise
determinations of the critical point, namely
$\kappa_c=0.074833(17)$ for $\gamma=1$ and
$\kappa_c=0.074819(15)$ for $\gamma=2$, in good
agreement with the value $\kappa_c=0.074834(15)$ obtained by Gaunt et al.
\cite{Gaunt:1979aa} from the symmetric phase.

In this sense, we can agree with Balog et al.:
a definitive test to decide between the two asymptotic behaviours
$\gamma=1$ and $\gamma=2$
has to be postponed to simulations performed
closer to the critical point where the non-leading terms associated with the
scale of the logs should become unessential.

However, although one can certainly get a good fit with $\gamma=1$,
the agreement between lattice data and the
prediction based on 2-loop renormalized perturbation theory is {\it not} good.
This can be checked considering the expression
reported by Balog et al. ($l=|\ln (\kappa -\kappa_c)|$)
\begin{equation}
\label{2loop}
      \left[ v^2_B \chi_2(0) \right]_{\text{2-loop}}=
a_1( {l} - \frac{25}{27} \ln { l}) +a_2
\end{equation}
together with the theoretical relations
\cite{Balog:2004zd}
\begin{equation}
\label{c2}
a_1=\frac{9(C'_2)^2}{32\pi^2}
\end{equation}
and
\begin{equation}
\label{cc2}
\frac{a_2}{a_1}=\ln(C'_3)+2\ln(C'_1)-1.6317
\end{equation}
Using the input values reported by L\"uscher and Weisz (LW) in Table 1 of
Ref~.\cite{Luscher:1988ek} and the relations $C'_1=e^{1/6}C_1$,
$C'_2=C_2$ and $C'_3=C_3$ (see Eqs.(4.37) and (4.38) of
Ref.\cite{Luscher:1988ek}), the predictions for the Ising model
($\bar{\lambda}=1$) are:
$C'_2=6.49(7)$, $2\ln(C'_1)=1/3+3.0(4)$ and $\ln(C'_3)=-3.0(1)$ or
$a_1=1.20(3)$, $a_2=-1.6(5)$.
However, the data for $v^2_B \chi_2(0)$ require
$a_1=1.267(14)$ and $a_2=-2.89(8)$, see Ref.\cite{Balog:2004zd}. Therefore,
the quality of the 2-loop fit is poor (see Fig.1).

%
%
%
%
\FIGURE[ht]{\label{Fig1}
\includegraphics[width=0.9\textwidth,clip]{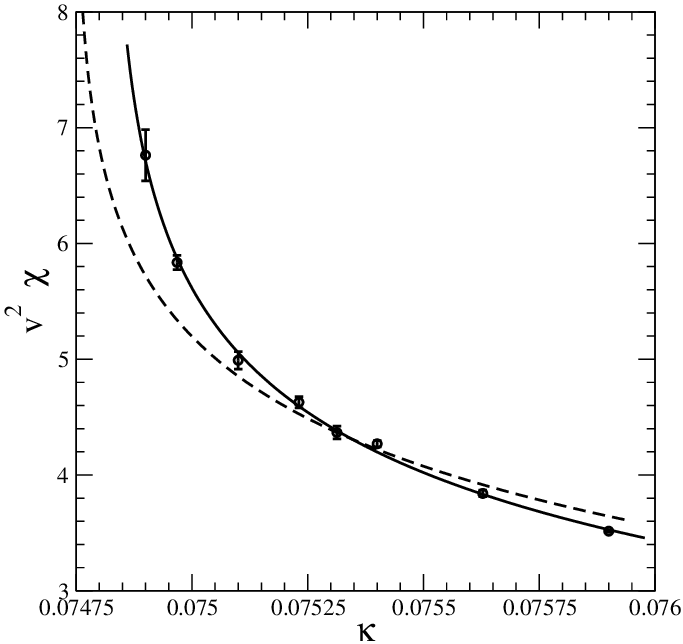}
\caption{
We show the lattice data for $v^2_B\chi_{\text{latt}}$
of Table 1
together with the fit Eq.~(\ref{Ffit}) for
$\gamma=2$ (solid line) and the two-loop fit Eq.~(\ref{2loop})
where the fit parameters
are allowed to vary inside their  theoretical uncertainties,
$a_1=1.20(3)$ and $a_2=-1.6(5)$ (dashed line).}}

Nevertheless, one can adopt a pragmatic point of view,
ignoring possible problems related to the matching conditions with the
symmetric phase, and try to extract from the data
for $v^2_B \chi_2(0)$ a new set of constants $C'_i$. In particular
the precise determination from $a_1=1.267(14)$ implies
\begin{equation}
\label{ccc2}
C'_2=6.67(4)
\end{equation}
that will be used in the following.

Thus we can summarize the results of this section as follows:

~~~~~~~~~~~1) leaving out the scale within the logs as a free parameter
(and thus allowing effectively for the presence of next-to-leading corrections)
the lattice data for $v^2_B\chi_2(0)$ are unable to distinguish between the
powers $\gamma=1$ and $\gamma=2$. In this sense, a definitive test
has to be postponed to simulations performed
closer to the critical point where the non-leading terms associated with the
scale of the logs should become unessential.

~~~~~~~~~~~2) a 2-loop fit as in Eq.(\ref{2loop}), with $a_1$ and
$a_2$ deduced consistently from the LW tables, does {\it not} provide a good
description of the data (see Fig.1). However, within the effective
2-loop formula Eq.(\ref{2loop}), when $a_1$ and $a_2$ are left as free
parameters, one can obtain precise determinations of the
critical point $\kappa_c=0.074833(17)$ and of the integration constant
$C'_2=6.67(4)$. These will be used in the next section to check whether
the lattice observables are consistent
with the critical behaviour predicted by Renormalization Group
(RG) analysis.

\section{Lattice tests of the alternative interpretation of `triviality'.
Part 2.}
\label{latticetest2}

Additional numerical
evidences concerning the relative scaling of $m_H$ and $\chi_2(0)$ can be
obtained by comparing again with the predictions of perturbation theory.
To this end, we performed
in Ref.\cite{Cea:2004ka} a test of the logarithmic trend predicted in
Eq.(\ref{z2phi}) assuming as input
entries the full 3-loop values $m_{\text{input}}\equiv m_R$ reported
in the first column of Table~3
of Ref.~\cite{Luscher:1988ek} at the various
values of $\kappa$. These input
mass  values, to leading order, follow the scaling law
\begin{equation}
\label{mrtau}
m_H =
A  \sqrt{\kappa -\kappa_c}\cdot |\ln(\kappa-\kappa_c)|^{-1/6}
\end{equation}
so that one can check whether the quantity
\begin{equation}
\label{zphi}
  Z_\varphi\equiv     2\kappa m^2_{\text{input}} \chi_{\text{latt}}
\end{equation}
tends to unity or grows logarithmically
when approaching the continuum limit.
\TABLE[t]{
\begin{tabular}{ccccccc}
$m_{\text{input}}$   &$\kappa$   &lattice  &algorithm &Ksweeps
&$\chi_{\text{latt}}$  &$v_B=\langle |\phi| \rangle$\\
\hline
0.4     &0.0759     &$32^4$  &S-W  & 1750  &41.714 (0.132)    & 0.290301 (21)                \\
0.4     &0.0759     &$44^4$  &W    &   60  &41.948 (0.927)    & 0.290283 (52)   \\ \hline
0.35    &0.075628   &$48^4$  &W    &  130  &58.699 (0.420)    & 0.255800 (18)   \\ \hline
0.3     &0.0754     &$32^4$  &S-W  &  345  &87.449 (0.758)    & 0.220540 (75)                \\
0.3     &0.0754     &$48^4$  &W    &  406  &87.821 (0.555)    & 0.220482 (19)   \\ \hline
0.275   &0.075313   &$48^4$  &W    &   53  &104.156 (1.305)   & 0.204771 (40)   \\ \hline
0.25    &0.075231   &$60^4$  &W    &   42  &130.798 (1.369)   & 0.188119 (31)   \\ \hline
0.2     &0.0751     &$48^4$  &W    &   27  &203.828 (3.058)   & 0.156649 (103)                \\
0.2     &0.0751     &$52^4$  &W    &   48  &201.191 (6.140)   & 0.156535 (65)                \\
0.2     &0.0751     &$60^4$  &W    &    7  &202.398 (8.614)   & 0.156476 (148)   \\ \hline
0.15    &0.074968   &$68^4$  &W    &   25  &460.199 (4.884)   & 0.112611 (51)   \\ \hline
0.1     &0.0749     &$68^4$  &W    &   24  &1125.444 (36.365) & 0.077358 (123)                \\
0.1     &0.0749     &$72^4$  &W    &    8  &1140.880 (39.025) & 0.077515 (210)  \\
\end{tabular}
\caption{
The details of the lattice simulations for each $\kappa$ corresponding to $m_{\text{input}}$.
In the algorithm column, 'S-W' stands for the Swendsen-Wang algorithm~\cite{Swendsen:1987ce},
while 'W' stands for the Wolff algorithm~\cite{Wolff:1989uh}. 'Ksweeps' stands for
sweeps multiplied by $10^3$.}
\label{Table2}
}
%
%
%

%
%
%
\FIGURE[ht]{\label{Fig2}
\includegraphics[width=0.9\textwidth,clip]{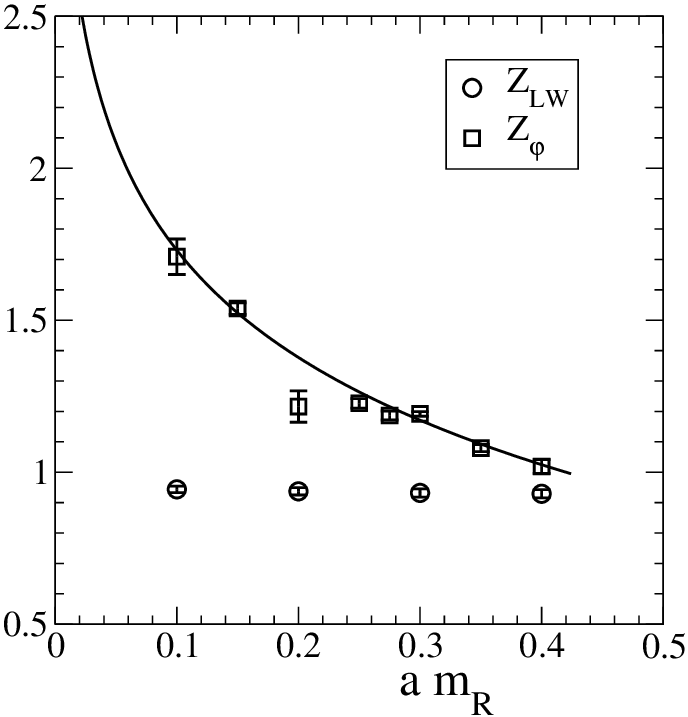}
\caption{
The lattice data for $Z_\varphi$, as defined in
Eq.~(\ref{zphi}), and its perturbative prediction $Z_{\text{LW}}$
versus the input mass values $m_{\text{input}}=a m_R$ taken from
the LW Table. The solid line is a fit to the form
$B \ln ( \Lambda/m_R)$ with $ \Lambda = \pi/a$ and $B=0.50$.
}}
%
%
%
%

%
%
%
%
\FIGURE[ht]{\label{Fig3}
\includegraphics[width=0.9\textwidth,clip]{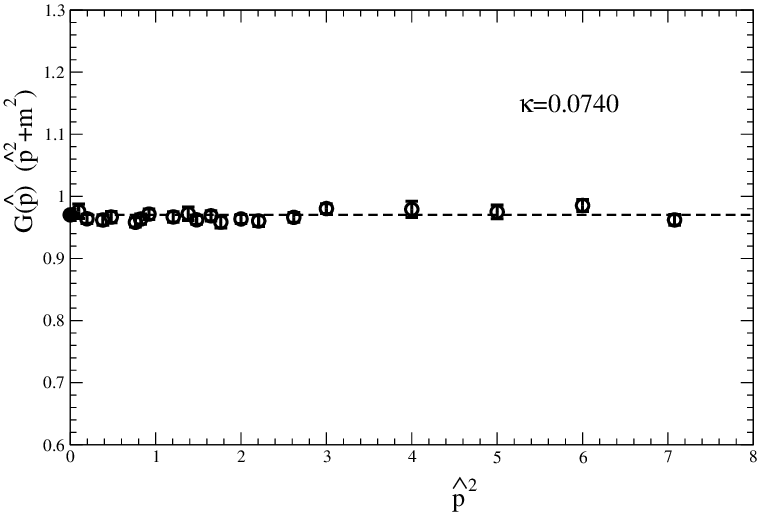}
\caption{
The lattice data for the re-scaled propagator in the symmetric phase at $\kappa=0.074$
as a function of the lattice momentum $\hat{p}_\mu =2 \sin p_\mu/2$.
The zero-momentum full point is defined as $Z_\varphi=(2\kappa)m^2_{\text{latt}}\chi_{\text{latt}}$.
The dashed line indicates the value of $Z_{\text{prop}}$.
Data are taken from Ref.~\cite{Cea:1999kn}.
}}
%
%
%
%
\FIGURE[ht]{\label{Fig4}
\includegraphics[width=0.9\textwidth,clip]{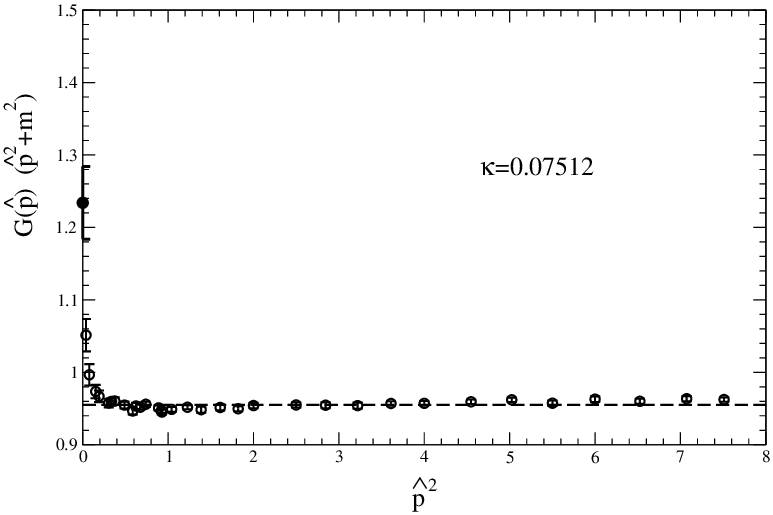}
\caption{
The lattice data for the re-scaled propagator in the broken phase at $\kappa=0.07512$
as a function of the lattice momentum $\hat{p}_\mu =2 \sin p_\mu/2$.
The zero-momentum full point is defined as $Z_\varphi=(2\kappa)m^2_{\text{latt}}\chi_{\text{latt}}$.
The dashed line indicates the value of $Z_{\text{prop}}$.
Data are taken from Ref.~\cite{Cea:1999kn}.
}}
%
%
%
%
%
%
%
%
\FIGURE[ht]{\label{Fig5}
\includegraphics[width=0.9\textwidth,clip]{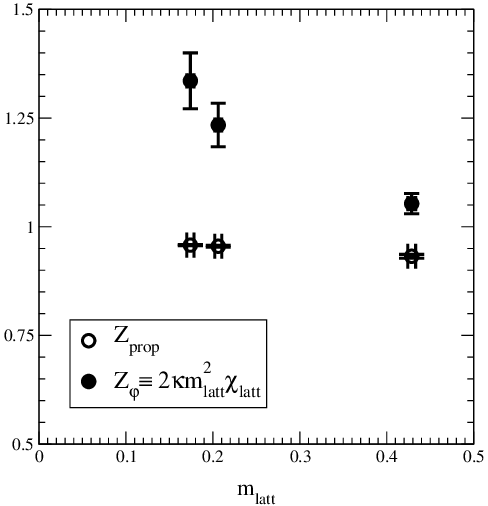}
\caption{
The two rescaling factors $Z_{\text{prop}}$ and
$Z_{\phi}$ obtained in Ref.~\cite{Cea:1999kn}  
as a function of $m_{\text{latt}}$.
}}
%
%
%
%
%
%
%
%
\FIGURE[ht]{\label{Fig6}
\includegraphics[width=0.9\textwidth,clip]{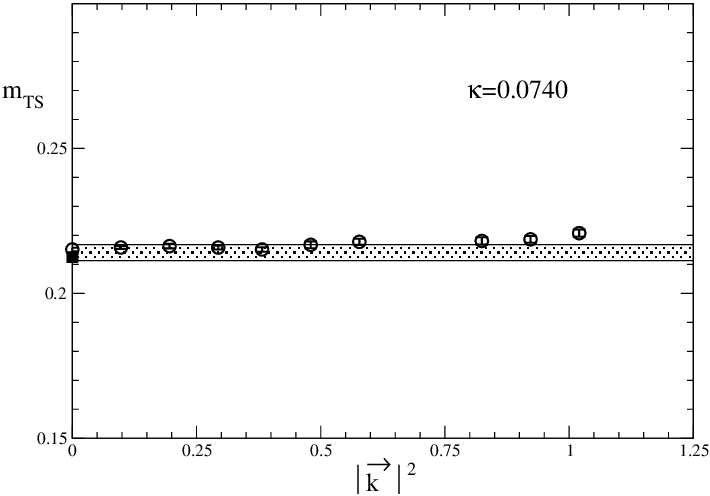}
\caption{
The time-slice mass $m_{\text{TS}}({\mathbf{k}})$ at different values of the three-momentum 
in the symmetric phase.
The shaded area represents the value $m_{\text{latt}}=0.2141(28)$ obtained
from the fit to the propagator data in Fig.3. The black square is
the result of Ref.~\cite{Montvay:1987us}.
}}
%
%
%
%
%
%
%
%
%
\FIGURE[ht]{\label{Fig7}
\includegraphics[width=0.9\textwidth,clip]{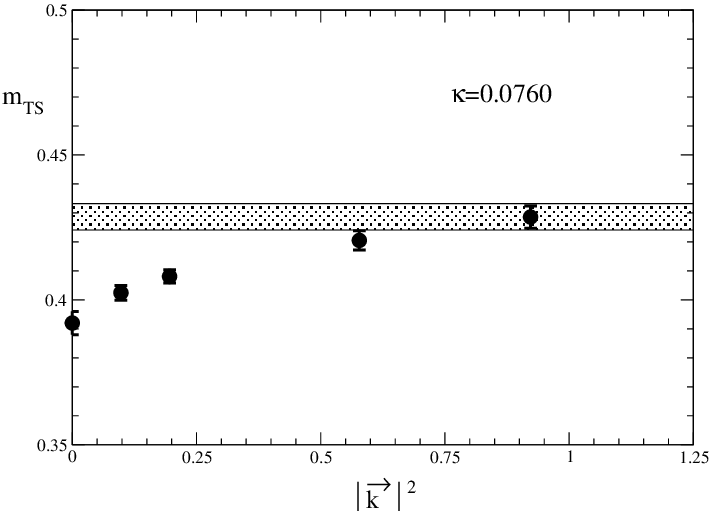}
\caption{
The time-slice mass $m_{\text{TS}}({\mathbf{k}})$
for several values of the spatial
momentum in the broken phase. The zero-momentum value is our result
$0.392(4)$ obtained from a simulation of 6 Msweeps
on a $20^4$ lattice (in perfect agreement with the value $0.392(1)$  reported in Table 4 of
Ref.\cite{Balog:2004zd}). The shaded area represents the value
$m_{\text{latt}}=0.42865(456)$ obtained from the fit
to the higher momentum propagator data.
}}

Now, using in Eq.(\ref{zphi}) the central values of
$m_{\text{input}}\equiv m_R$ reported in the LW Table and
the values of $\chi_{\text{latt}}$ reported
in Table 1, the conclusion is unambiguous: the
$Z_\varphi$ in Eq.(\ref{zphi})
becomes larger and larger
approaching the continuum limit along the RG curve
$m_{\text{input}}=m_{\text{input}}(\kappa)$ and
the observed increase is completely consistent with the
logarithmic trend predicted in Eq.(\ref{z2phi}) (see Fig.2).

The discrepancy with the perturbative predictions
can also be checked noticing that for $m_{\rm input}=0.08$
Ref.~\cite{Luscher:1988ek} predicts $\kappa=0.07481(8)$, i.e.
{\em smaller} than $0.0749$. Therefore, by inspection
of Table 1, the relevant
lattice susceptibility will be
definitely {\em larger} than its value for $\kappa=0.0749$,
$\chi_{\rm latt}\sim 1100$, so that
using Eq.~(\ref{zphi}), one gets the {\em lower bound}
$Z_\varphi > 1.05$ which
cannot be reconciled with the  perturbative predictions.

Balog et al. object to our conclusions that
"..the crucial question is whether the estimates
$m_{\text{input}}$ of $m_R$ are reliable". According to these authors,
"..the measured values of $m_R$ are considerably lower than the corresponding
estimates $m_{\rm input}$" (i.e. the $m_R$'s  that we took
from the LW table). As a matter of fact,
by replacing the LW $m_R$'s with the results of their
simulations, one gets a remarkably constant value
$2\kappa m^2_R\chi_{\text{latt}}\sim 0.88$.

Thus their objection
does not concern our strategy but rather the validity of the
LW entries themselves as reliable
estimates of the Higgs mass parameter $m_H$.
Actually, as we shall illustrate in the following,
their conclusion does not apply: to check the true behaviour of
$Z_\varphi=2\kappa m^2_H\chi_{\text{latt}}$
one should first identify correctly the operative definition of
$m_H$ on the lattice. The mass values reported by Balog et al. are not
{\it reliable} determinations of the physical Higgs mass
if one requires the theoretical
consistency of the adopted definition of `mass'.
To fully appreciate what is going on, we need to go back to
Refs.~\cite{Cea:1998hy,Cea:1999kn,Cea:1999zu} (otherwise
ignored by the authors of Ref.\cite{Balog:2004zd}).

In those calculations, one was fitting the lattice data
for the connected propagator to the (lattice version of the)
two-parameter form
\begin{equation}
\label{gprop}
G_{\text{fit}}(p)= \frac{Z_{\text{prop}}}{ p^2 + m^2_{\text{latt}} }  \,.
\end{equation}
This is a clean and simple strategy:
there is no reason to restrict the analysis of the propagator
to the limit $p\to 0$. In fact,
in a free-field theory (the lattice version of)
Eq.(\ref{gprop}) is valid in the full
range $0\leq p^2 \leq \Lambda^2$ (with $\Lambda\sim \pi/a$). In a `trivial'
theory  one expects a two-parameter fit to the propagator data to have small
residual corrections. These,
however, should become smaller and smaller by
approaching the continuum limit.

In this way, after computing the lattice zero-momentum susceptibility
$\chi_{\text{latt}}$, it becomes possible to compare the measured value of
$Z_\varphi \equiv 2\kappa m^2_{\text{latt}} \chi_{\text{latt}}$
with the fitted $Z_{\text{prop}}$, both in the symmetric and broken phases.
While no difference was found in the symmetric phase (see Fig.~3),
$Z_\varphi$ and $Z_{\text{prop}}$ were found to be sizeably different
in the broken phase. The discrepancy was found to become larger and larger
by approaching the continuum limit and thus cannot be explained in terms
of residual perturbative corrections to Eq.(\ref{gprop}).
In particular, $Z_{\text{prop}}$ was very slowly
varying and steadily approaching unity from below in the continuum limit
consistently with K\'allen-Lehmann representation and `triviality'.
$Z_\varphi$, on the other hand,
was found to rapidly increase {\em above} unity in the same limit. The
observed trend was consistent with the logarithmically increasing
trend predicted in Eq.(\ref{z2phi}).

This conclusion was based on the values of $m_{\text{latt}}$
extracted by skipping the lowest 3-4 $p^2-$values in the fit to
the propagator data. In fact, differently from the symmetric-phase
simulations, where Eq.(\ref{gprop}) reproduces the data in the full
range $0\leq p^2 \leq \Lambda^2$,
in the broken-symmetry phase the two-parameter form
Eq.~(\ref{gprop}) does not reproduce the
propagator data down to $p=0$, see Fig.4
(and Figs.3,5,6 of Ref.\cite{Cea:1999kn}).
Thus one has to choose. Either a) to restrict to
the lowest 3-4 $p^2-$ data, and obtain from the fit a pair of values
$(Z_{\text{low}},m_{\text{low}})$ or b) skip these first
few data and fit the much
larger sample of higher momentum data thus obtaining the other pair
$(Z_{\text{high}},m_{\text{high}})$.

The two sets of mass values differ sizeably.
For instance for
$\kappa=0.076,0.07512,0.07504$ the results of
Refs.~\cite{Cea:1999kn} were respectively
$m_{\text{high}}=0.4286(46),0.2062(41),0.1723(34)$. On the other hand, the
alternative values from the lowest momentum data are
respectively $m_{\text{low}}=
0.392(4),0.1737(24),0.1419(17)$ for the same $\kappa$'s.

Numerically, the typical $m_{\text{low}}$'s are extremely
close to the other mass definition $m_{\text{TS}}({\bf{k=0}})$,
the mass extracted at zero 3-momentum from the exponential decay
(TS=`Time Slice') of the connected two-point correlator
\begin{equation}
\label{corr}
C_1(t,0; {\bf k})\equiv \langle
S_c(t;{\bf k})S_c(0;{\bf k})+
S_s(t;{\bf k})S_s(0;{\bf k}) \rangle _{\rm conn} ,
\end{equation}
where
\begin{equation}
\label{cos}
S_c(t; {\bf k})\equiv \frac{1}{L^3} \sum _{ { \bf x} } \phi({\bf x}, t)
\cos ({\bf k} \cdot {\bf x}) ,
\end{equation}
\begin{equation}
\label{sin}
S_s(t;{ \bf k})\equiv \frac{1}{L^3} \sum _ {{\bf x}} \phi({\bf x}, t)
\sin ({\bf k} \cdot {\bf x}) .
\end{equation}
Here, $t$ is the Euclidean time; ${\bf x}$ is the spatial part of the site
4-vector $x^{\mu}$; ${\bf k}$ is the lattice momentum
${\bf k}=(2\pi/L) (n_x,n_y,n_z$), with $(n_x,n_y,n_z)$ non-negative integers;
and $\langle ...\rangle_{\rm conn}$ denotes the connected expectation
value with respect to the lattice action, Eq.~(\ref{ising}). In this way,
parameterizing the correlator $C_1$ in terms of the energy $E_k$ as
($L_t$ being the lattice size in time direction)
\begin{equation}
\label{fitcor}
C_1(t,0;{\bf k})= A \, [ \, \exp(-E_k t)+\exp(-E_k(L_t-t)) \, ] \,,
\end{equation}
the mass can be determined through the lattice dispersion relation
\begin{equation}
\label{disp}
m^2_{\rm TS}({\bf k}) = ~2 (\cosh E_k  -1)~~ -~~2 \sum ^{3} _{\mu=1}~
(1-\cos k_\mu) \,.
\end{equation}
In a free-field theory $m_{\rm TS}$ is independent of {\bf k} and
coincides with $m_{\text{latt}}$ from Eq.~(\ref{gprop}).

Now, using the corresponding susceptibility values
$\chi_{\text{latt}}=$37.85(6), 193.1(1.7), 293.4(2.9) reported in
Ref.~\cite{Cea:1999kn} and the above values of
$m_H=m_{\text{high}}$ one obtains a set of logarithmically
increasing values (see Fig.5):
$Z_\varphi \equiv (2\kappa) m^2_H \chi_{\text{latt}}=
1.05(2),1.23(5),1.31(5)$ as predicted in
Refs.\cite{Consoli:1994jr,Consoli:1997ra,Consoli:1999ni}.

This is in contrast with the values of the other quantity
$Z_{\text{low}}\sim
\hat{Z}_R\equiv (2\kappa) m^2_{\text{low}} \chi_{\text{latt}}=
0.884(18),0.875(24),0.887(21)$ which remain remarkably stable
(see the
corresponding entries for $\hat{Z}_R$ shown in Table 4 of
Ref.\cite{Balog:2004zd}).

Therefore, the crucial question raised by the propagator data
is the following. Which is
the `true' lattice definition of $m_H$ ? Is the $p\to 0$
choice (adopted by Balog et al.) and represented by
$m_{\text{low}}\sim m_{\text{TS}}({\mathbf{k=0}})$,
or that obtained from
$m_{\text{high}}$, as proposed in
Refs.~\cite{Cea:1999kn}?
As discussed in Ref.\cite{Cea:1999kn}, there are several arguments that suggest
the correct definition of $m_H$ to be obtained from
$m_{\text{high}}$, thus regarding
the other values extracted from the very low-momentum
region as a symptom of the distinct dynamics
of the scalar condensate. Here we list a few:

~~~~~~~~~~~~i) in the continuum theory, the shifted fluctuation field
is defined as the $p_\mu\neq 0 $ projection of the full quantum field.
However in a lattice simulation with periodic boundary conditions, where the
momenta $p_i$ are proportional to an integer number $n_i$ times
the inverse lattice size $1/L$, the
notion $p_\mu\neq 0$ is ambiguous. In fact, {\it any} finite set of
integers will evolve onto the $p_\mu=0$ state in the limit $L \to \infty$.
This means that, for a given lattice mass,
by increasing the lattice size, to separate unambiguously
the genuine
finite-momentum fluctuation field from the `condensate' itself,
one should increase correspondingly the set of integers
$n_i$. This gives a
clean physical meaning to the fit obtained by skipping the lowest momentum
propagator data and to the pair of parameters
$(Z_{\text{high}},m_{\text{high}})$.

~~~~~~~~~~~~ii) the values obtained in
Ref.~\cite{Cea:1999kn}
(respectively for $\kappa=$0.076, 0.07512,0.07504)
$m_{\text{high}}=$0.4286(46), 0.2062(41), 0.1723(34)
give a physical mass $m_H$ that scales
as expected. This can easily be checked from the
remarkable agreement between these values and
the predicted trend Eq.(\ref{mrtau}) which is valid at the leading-log
level both in perturbation theory and in the alternative picture of
Refs.\cite{Consoli:1994jr,Consoli:1997ra,Consoli:1999ni},
see Eqs.(\ref{ltriv}) and (\ref{lambdaR2}).
In this way one finds $A=17.282(364)$, $\kappa_c=0.074836(15)$
and predicts $m_H\sim$0.499, 0.408, 0.293, 0.199
for $\kappa=$0.0764, 0.0759, 0.0754, 0.0751 in good agreement
with the values $0.5, 0.4, 0.3, 0.2$ reported in the LW Table.

~~~~~~~~~~iii) the identification of
$m_{\text{low}}\sim m_{\text{TS}}({\mathbf{k=0}})$ as the physical $m_H$
contradicts the observed
dependence of $m_{\text{TS}}({\mathbf{k}})$ on
$|{\mathbf{k}}|$ in the limit
${\mathbf{k}}\to 0$. In fact, in the symmetric phase 
(see Fig.~6) the energy spectrum  has the form
$\sqrt{ {\mathbf{k}}^2+ m^2 }$ up to momenta
${\mathbf{k}}^2 \sim 25 m^2$(!).
However in the broken phase the energy is not reproduced by the 
(lattice version of the)
form $\sqrt{ {\mathbf{k}}^2+ {\text{const}} }$ (see Fig.7) and
thus the very notion of `mass' becomes problematic.
At larger ${\mathbf{k}}$, where
$m_{\text{TS}}({\mathbf{k}})$ becomes insensitive to ${\mathbf{k}}$, 
it agrees well with $m_{\text{high}}$.
The very different behaviour between symmetric and broken phase shown
in Figs.~6 and 7 has no counterpart in the conventional picture. 

~~~~~~~~~~~iv) the values of
$Z_{\text{prop}}$ obtained from the fit to the higher-momentum data in
Ref.~\cite{Cea:1999kn} for $\kappa=0.076,0.07512,0.07504$,
$Z_{\text{high}}=$0.9321(44), 0.9551(21), 0.9566(13),
exhibit a monotonical increase toward
unity (from below) as expected on the base of
the K\'allen-Lehmann representation
in a `trivial' theory. This confirms that fitting the propagator skipping the
lowest momentum data
gives consistent results. On the contrary, the values for
$Z_{\text{low}}\sim \hat{Z}_R$
obtained from the fit to the lowest momentum points alone, remain
constant to $\sim 0.88$
in the limit $\kappa \to \kappa_c$ (see also the values in
Table 4 of Ref.\cite{Balog:2004zd}). Assuming this definition as the correct
one to be used in
the K\"allen-Lehmann representation, one would find
an inner contradiction with the
non-interacting nature of the shifted fluctuation field in the continuum limit.

No trace of this discussion is found in
Ref.\cite{Balog:2004zd}. These authors, ignoring the evident difference
between Fig.3 and Fig.4, as well as between Fig.6 and Fig.7,
do not address the physical consistency of the mass
definition extracted from the very low-momentum region of the lattice data.
They just limit themselves to the remark
that, for the $\lambda\Phi^4$ case,
the lattice propagator is consistently reproduced by
Eq.(\ref{gprop}) for $p \to 0$. However, this agreement is
not relevant here since the Ising limit is known to
anticipate much better the true aspects of the theory
in finite lattices. In the Ising case, in fact, the fit is not good
and Balog
et al. are forced to extract the lattice mass by fitting just
to the three lowest momentum points, precisely as with the parameter
$m_{\text{low}}$ of Ref.~\cite{Cea:1999kn}.

It is also surprising their claim (see the Abstract of
Ref.\cite{Balog:2004zd}) that the lattice data are consistent with the
perturbative predictions. We have seen in the previous section
that, if one requires consistency between
2-loop predictions and the data for $v^2_B\chi_2(0)$, one has to replace the
integration constant $C'_2$ given in the LW's Table 1 with the value
$C'_2=6.67(4)$ given in Eq.(\ref{ccc2}). Therefore, using the perturbative
relation reported in Eq.(2.14) of Ref.\cite{Balog:2004zd}, namely
($\alpha_R=\frac{g_R}{16\pi^2}$)
\begin{equation}
\label{ZRhat}
\hat{Z}_R=(2\kappa)C'_2 ( 1- \frac{7}{36} \alpha_R + {\cal O}(\alpha^2_R))
\end{equation}
one can compare this $\hat{Z}_R$ with the other estimate
extracted from the product
$m^2_R\chi_2(0)$ and reported in Table 4 of
Ref.\cite{Balog:2004zd}.

In this case, for $\kappa=0.076$
(where the relevant value is $g_R= 30.37(28)$) Eq.(\ref{ZRhat}) predicts
$\hat{Z}_R= 0.976(6)$ while the value reported in Table 4 by Balog et al. is
$\hat{Z}_R= 0.896(4)$. Analogously,
for $\kappa=0.0751$
(where the relevant values is $g_R= 20.51(74)$) Eq.(\ref{ZRhat}) predicts
again $\hat{Z}_R= 0.976(6)$
while the value reported in Table 4 by Balog et al. is
$\hat{Z}_R= 0.883(17)$. These discrepancies, that are at the level of
$\sim 10\sigma$ and $\sim 5\sigma$, can hardly be considered indicative
of theoretical consistency.

Quite independently, the authors of
Ref.\cite{Balog:2004zd} have not shown that the
$m_R$'s reported in their Table 4 lie on well defined RG trajectories
$m_R=m_R(\kappa)$ using the {\it same}
$\kappa_c=0.074833(17)$ obtained
from the 2-loop fit to the data for $v^2_B\chi_2(0)$.
In particular, this means that
their value $m_R=0.395(1)$ for $\kappa=0.076$
has to come out
consistent with the other value $m_R=0.1688(15)$ for $\kappa=0.0751$.
To clarify this issue, they should quote the new values of
the mass that replace the LW entries for all values of $\kappa$.

\section{Summary and conclusions}
\label{summary}

Let us now try to summarize the various
theoretical and numerical points addressed
in this paper. We shall start by observing that
the field rescaling is usually viewed
as an `operatorial statement' between bare and renormalized fields
operators of the type
\begin{equation}
\label{operator}
                     "~\Phi_B(x)= \sqrt{Z} \Phi_R(x)~"
\end{equation}
As pointed out in Ref.\cite{Agodi:1995qv},
this relation is a consistent short-hand notation
in a theory where the field operator admits an
asymptotic Fock representation, as in QED. In the presence of
spontaneous symmetry breaking it has no rigorous
basis since the Fock representation exists only for the {\em shifted}
fluctuating field $h_B(x)= \Phi_B(x) -\langle \Phi_B \rangle$, the one
with a vanishing expectation value.

For this reason, following
Refs.\cite{Consoli:1994jr,Consoli:1997ra,Consoli:1999ni}
(see the discussion given in Section 2), one can consider
theoretical frameworks that are fully consistent with `triviality' but
where such an operatorial relation is {\it not} valid. In fact,
the physical conditions used to determine the
$h_B-h_R$ relation, through a $Z=Z_{\text {prop}}\sim 1$ in Eq.(\ref{hr}),
can be basically different from those used in the
$v_B-v_R$ case, through a $Z=Z_\varphi\sim \ln \Lambda$ in Eq.(\ref{vR}).
In this case, if one wants to introduce a renormalized field operator
$\Phi_R(x)$ (whose vacuum expectation value is what one defines by
$v_R$ and whose fluctuating part
is what one defines by $h_R$), this cannot
be related to $\Phi_B(x)$ by means of Eq.(\ref{operator}).

As reviewed at the end of Section 2,
this `subtlety' has a substantial phenomenological implication: `triviality',
by itself, cannot be used to place upper bounds on the Higgs boson mass.
Thus, the importance of the issue
requires dedicated lattice simulations
to check the validity of the logarithmic trend predicted in Eq.(\ref{z2phi}) by
measuring the relative scaling of the physical Higgs mass squared $m^2_H$
vs. the zero-momentum susceptibility.

Clearly, the answer to this question
depends on the given procedure adopted on the lattice
to extract $m_H=m_{\text{latt}}$. Our point is that a correct choice
requires to fulfill several consistency tests, taking into account the
free-field nature of the fluctuation field in the continuum limit.
Thus one should first
find a 4-momentum region where the connected propagator
is described by the (lattice version of the) two-parameter form
\begin{equation}
G_{\text{fit}}(p)= \frac{Z_{\text{prop}}}{ p^2 + m^2_{\text{latt}} }  \,.
\end{equation}
One should also check that the fitted value
$m_H=m_{\text{latt}}$ controls the energy eigenvalues
$E=E({\mathbf{k}})$ governing the exponential time
decay of the connected correlator in a given region of 3-momentum ${\mathbf{k}}$.
In fact, where the energy is not reproduced by the (lattice version of the)
form $\sqrt{ {\mathbf{k}}^2 + { \text{const.}} }$, the very notion of mass
becomes problematic. Finally, one should also check that the fitted values of
$Z_{\text{prop}}$ approach unity from below in the continuum limit as required
by the K\'allen-Lehmann representation in a `trivial' theory.

Now, the usual assumption is to extract
$m_H=m_{\text{latt}}$ from the
$p \to 0$ and/or ${\mathbf{k}} \to 0$ limits. This is certainly
valid in a simulation performed in the symmetric phase (see Figs.3 and 6)
where the whole 3- and 4-momentum regions
give the same indications.
However, in the broken-symmetry phase, where
the vacuum is some sort of `condensate',
there might be reasons to avoid the strict zero-momentum limit to
extract the physical particle mass.

For instance, as mentioned in Sect.4,
in the continuum theory, the shifted fluctuation field
is defined as the $p_\mu\neq 0 $ projection of the full quantum field.
However in a lattice simulation with periodic boundary conditions, where the
momenta $p_i$ are proportional to an integer number $n_i$ times
the inverse lattice size $1/L$, the
notion $p_\mu\neq 0$ is ambiguous. In fact, {\it any} finite set of
integers will evolve onto the $p_\mu=0$ state in the limit $L \to \infty$.
Therefore, for a given lattice mass,
by increasing the lattice size, to separate unambiguously
the genuine
finite-momentum fluctuation field from the `condensate' itself,
one should increase correspondingly the set of integers $n_i$.

In addition, there are precise physical motivations suggested by
the non-relativistic limit of a broken-symmetry $\lambda\Phi^4$
theory: the low-temperature phase of a
hard-sphere Bose gas \cite{Huang:1957}. In this case, the low-lying
excitations for ${\mathbf{k}}\to 0$
are phonons, i.e. collective oscillations of the hard-sphere
system whose energy grows linearly
$E_{\text{ph}}({\mathbf{k}})\sim c_s |{\mathbf{k}}|$, $c_s$ being the speed of sound.
Only at {\it larger} $|{\mathbf{k}}|$ does the energy spectrum grow quadratically.
Therefore, a determination of the effective hard-sphere mass
through the non-relativistic 1-particle relation
$E_{\text{1-part}}({\mathbf{k}})\sim \frac { {\mathbf{k}}^2 }{2 m_{\text{eff}}} $ cannot
be obtained from the ${\mathbf{k}} \to 0$ limit of the energy spectrum which is
dominated by the phonon branch.

This type of problems were preliminarily considered in
Ref.~\cite{Cea:1999kn}. The result of that investigation
was that the physical mass $m_H=m_{\text{latt}}$ is obtained from the propagator data
after skipping the lowest 3-4 momentum points.
The mass value $m_{\text{low}}$, obtained from the lowest momentum points, 
which is numerically close to the other
mass extracted from the exponential decay of the connected correlator
at zero 3-momentum, does not fulfill the same consistency checks.

Now, for $\kappa=0.076,0.07512,0.07504$ the results of
Ref.~\cite{Cea:1999kn} were respectively
$m_H=m_{\text{latt}}=0.4286(46),0.2062(41),0.1723(34)$. In this way,
using the susceptibility values
$\chi_{\text{latt}}=$37.85(6), 193.1(1.7), 293.4(2.9) reported in
Ref.~\cite{Cea:1999kn} one obtains a set of logarithmically
increasing values:
$Z_\varphi \equiv (2\kappa) m^2_H \chi_{\text{latt}}=
1.05(2),1.23(5),1.31(5)$ as predicted in
Refs.\cite{Consoli:1994jr,Consoli:1997ra,Consoli:1999ni} (see Fig.5).
This is in contrast with the values obtained using $m_{\text{low}}$ for which
the quantity
$\hat{Z}_R\equiv (2\kappa) m^2_{\text{low}} \chi_{\text{latt}}\sim
0.88$ remains remarkably stable.

To provide {\it further} evidence, we replaced in
Ref.\cite{Cea:2004ka} the direct evaluation of
$m_H=m_{\text{latt}}$ with the theoretical input values predicted in the
LW Tables. This is not in contradiction with the previous strategy, in fact
the values obtained in Ref.~\cite{Cea:1999kn}
$m_{\text{latt}}=$0.4286(46), 0.2062(41), 0.1723(34)
give a physical mass $m_H$ that scales as in
Eq.(\ref{mrtau}) and that  is
in good agreement
with the values $m_H=m_{\text{input}}(\kappa) $ reported in the LW Table.
Using the values of $\chi_{\text{latt}}$
reported in our Table 1 and the LW entries for the mass,
this additional test confirms that the quantity
  $Z_\varphi\equiv     2\kappa m^2_{\text{input}} \chi_{\text{latt}}$
increases logarithmically when approaching the continuum limit (see Fig.2).

Now, Balog et al., being aware that the above numerical evidences
have "..serious non standard implications for the Higgs sector of the
Standard Model" (see the Conclusions of Ref.\cite{Balog:2004zd}), have
performed a new analysis. They claim that the LW entries for $m_R$
should be replaced by new values (whose consistency with RG trajectories
$m_R=m_R(\kappa)$, however, has not been shown). As far as we can see,
they have essentially re-discovered the result of
Ref.~\cite{Cea:1999kn} that the quantity
$\hat{Z}_R\equiv (2\kappa) m^2_{\text{low}} \chi_{\text{latt}}$
is a constant $\sim 0.88$. However, ignoring
Ref.~\cite{Cea:1999kn}, they fail to appreciate why their values
$m_R=m_{\text{low}}$ do not represent a consistent
lattice definition of $m_H$.

A point where we accept their criticism concerns the lattice data
for $v^2_B \chi_2(0)$. Restricting to this observable,
a definitive test of the leading-logarithmic trend
has to be postponed to data taken closer to the critical point where the
non-leading terms associated with the scale of the logs should become
unessential.

However, as discussed at the end of Sect.3, within the perturbative
framework, the data for $v^2_B \chi_2(0)$ give precise information
on the integration constants $C'_i$ (i=1,2,3) that should be
used for the matching with the symmetric phase.
Using the precise outcome of the fit $C'_2=6.67(4)$
Eq.~(\ref{ZRhat}) predicts
$\hat{Z}_R\sim 0.976(6)$ with 5-10$\sigma$ discrepancies with respect
to the values reported by Balog et al in their Table 4.

This is precisely the same discrepancy pointed out by Jansen et al.
Ref.\cite{Jansen:1989cw} whose origin cannot be understood
ignoring the main point of Ref.~\cite{Cea:1999kn} and of this paper.
In the broken phase, a naive zero 4-momentum limit of the connected propagator
and/or a naive zero 3-momentum limit of the energy eigenvalue
controlling the exponential decay of the connected correlator,
do not provide consistent estimates of the physical mass parameter $m_H$.
The phenomenological implications are substantial. Once the quadratic shape
of the effective potential (the inverse susceptibility), that is a pure
zero-momentum quantity, does not scale uniformly with $m^2_H$, although the
finite-momentum fluctuation field becomes free in the continuum limit, one
cannot use `triviality' to place upper bounds on $m_H$.

\providecommand{\href}[2]{#2}\begingroup\raggedright\endgroup

\end{document}